\documentclass[10pt,journal,compsoc]{IEEEtran}
\ifCLASSOPTIONcompsoc
  \usepackage[nocompress]{cite}
\else
  \usepackage{cite}
\fi
\ifCLASSINFOpdf
  \usepackage[pdftex]{graphicx}
  \DeclareGraphicsExtensions{.pdf,.jpeg,.png,.eps}
\else
\fi
\usepackage{color}
\usepackage{multirow, makecell, longtable}
\usepackage{amsmath}
\usepackage{times}
\usepackage{helvet}
\usepackage{multirow, makecell, longtable}
\usepackage{subfigure}
\usepackage{algorithmic}
\usepackage[ruled,vlined]{algorithm2e}
\usepackage{booktabs}
\usepackage{courier}

\begin{document}
\title{Time-aware Gradient Attack on Dynamic Network Link Prediction}
\author{Jinyin~Chen,
        Jian~Zhang,
        Zhi~Chen,
        Min~Du,
        and~Qi~Xuan,~\IEEEmembership{Member,~IEEE}
}

\maketitle

\begin{abstract}
In network link prediction, it is possible to hide a target link from being predicted with a small perturbation on network structure. This observation may be exploited in many real world scenarios, for example, to preserve privacy, or to exploit financial security. There have been many recent studies to generate adversarial examples to mislead deep learning models on graph data. However, none of the previous work has considered the dynamic nature of real-world systems. In this work, we present the first study of adversarial attack on \textit{dynamic network link prediction} (DNLP). The proposed attack method, namely \textit{time-aware gradient attack} (TGA), utilizes the gradient information generated by deep dynamic network embedding (DDNE) across different snapshots to rewire a few links, so as to make DDNE fail to predict target links. We implement TGA in two ways: one is based on traversal search, namely TGA-Tra; and the other is simplified with greedy search for efficiency, namely TGA-Gre. We conduct comprehensive experiments which show the outstanding performance of TGA in attacking DNLP algorithms.
\end{abstract}


\section{Introduction}\label{sec:introduction}
\IEEEPARstart{I}{n} the era of big data, network analysis emerges as a powerful tool in various areas, such as recommendation on e-commerce websites~\cite{gong2019exact}, bug detection in software engineering~\cite{zhang2016correlation}, and behavioral analysis in sociology~\cite{fionda2018community}. Existing algorithms like traditional similarity indices and newly developed network embedding methods learn structural features on static networks while most of them ignore the dynamic nature of real-world systems. The dynamics of networks, such as the establishment and dissolution of friendship in online social networks, could help the analysis. For example, in citation networks, the citations in different time which reflect the research focus of the authors should weigh differently in the prediction of future citations. And dynamic networks are naturally designed in these occasions.

Dynamic network analysis, especially link prediction~(DNLP) which we discuss in the paper, mainly focuses on the linkages status. That is predicting whether a link exists or not at a certain time based on historical information. For instance, given a social network, DNLP algorithms predict a given individual's relationships with others in the near future based on his/her historical friend lists. Compare with traditional link prediction algorithms, DNLP algorithms could learn the underlying transitions of behavior patterns while the traditional ones could not. DNLP is generally performed in two ways: time-series data analysis and machine learning based methods. A representative category of the former one is Autoregressive Integrated Moving Average~(ARIMA)~\cite{gunecs2016link, ozcan2015multivariate} related methods which deal with time varying similarity scores as time series to predict future links. The latter includes factorization based approaches and deep learning models of which the majority consist of Recurrent Neural Networks~(RNNs).

Regardless of their performance, DNLP algorithms may suffer from adversarial attacks as most machine learning methods do. The exploitation of adversarial examples may expose DNLP algorithms to security threats. However, from another perspective, this property could be useful in other fields like privacy-preserving. Privacy issues have aroused wide concern since the data theft of Facebook in 2018. In fact, one can infer private information, such as romantic relationship or visiting the same restaurant~\cite{fu2018link, xuan2018modern}, of target individuals with the help of advanced algorithms like DNLP. A well designed adversarial example may protect intimate relationships from being predicted by even the most advanced DNLP approach, which could provide a possible solution to privacy protection. The target link could be hidden by linking the user to someone unfamiliar, or removing intimate links in historical interactions. A bunch of works have explored the ways of generating adversarial examples for graphs in recent years. However, few of them involve dynamic networks, not even DNLP.

In this paper, we propose a novel adversarial attack targeting DNLP, which we refer as \emph{Time-aware Gradient Attack}~(TGA), to hide target links from being predicted. Benefitting from the gradients generated by deep learning model, i.e., DDNE, TGA is able to find candidate links to be modified without extensive search, and perform attack at minimum cost. Considering the dynamics of networks, TGA compares the gradients on different snapshots separately rather than does simple sorting on all snapshots; furthermore, it searches candidate links across iterations to make full use of the gradients.
Overall, our main contributions are summarized as follows.

\begin{itemize}
    \item We design TGA to generate adversarial examples based on the gradients obtained by DDNE. As far as we know, it is the first work about adversarial attacks on DNLP.
    \item We conduct extensive experiments on four real-world dynamic networks and compare TGA with several baselines. The results show that TGA achieves the state-of-the-art attack performance. And a case study on Ethereum transaction network is carried out to validate the practicability.
    \item We vary DNLP model parameters and observe several interesting phenomena which could be inspiring to future research. For example, long-term prediction is more vulnerable to adversarial attacks; while integrating more historical information can increase the robustness of DDNE.
\end{itemize}
For the rest of this paper, we first review related works in Sec.~\ref{sec:relate} and preliminaries in Sec.~\ref{sec:preliminary}. Then, we present the proposed attack details of TGA-Tra and TGA-Gre in Sec.~\ref{sec:method}, and gives the results of the proposed attack methods as well as the performance under some certain circumstances in Sec.~\ref{sec:experiment}. Finally, we conclude the paper with the prospect of future work in Sec.~\ref{sec:conclusion}.

\section{Related Work}
\label{sec:relate}
This section briefly reviews the literature of DNLP algorithms and the related work on adversarial attacks.

\textbf{DNLP algorithm.}
Recently, a temporal restricted Boltzmann machine (TRBM) was adopted with additional neighborhood information, namely ctRBM~\cite{li2014deep}, to learn the dynamics as well as the structural characteristics of networks. As an extension of ctRBM, GTRBM~\cite{li2018deep} combines TRBM and boosting decision tree to model the evolving pattern of each node. Besides the RBM-based methods, recurrent neural networks~(RNN), like long short-term memory~(LSTM), plays an important role in other DNLP algorithms. A stacked LSTM module was applied inside the autoencoder framework to capture time dependencies of the whole network~\cite{chen2019lstm}, and a gated recurrent network~(GRU) was used as the encoder which could relatively lower the computational complexity~\cite{li2018deep}. There are also many other methods based on random walk~\cite{nguyen2018continuous}, matrix factorization~\cite{zhang2018timers} and so forth~\cite{zhou2018dynamic}.

\textbf{Adversarial attack.}
A bunch of works have explored the field of adversarial attack on graph data. Community membership anonymization was realized by connecting the target user to the one of high centrality~\cite{nagaraja2010impact}.  Another method focused on disconnecting certain neighbors while adding links between different communities, with regards to the centrality of degree, closeness and betweenness~\cite{waniek2018hiding}. In fact, community deception can be achieved by only rewiring the links inner the target community~\cite{fionda2018community}. On the other hand, the emerging network embedding techniques, such as the graph convolutional network~(GCN)~\cite{kipf2016semi}, have drawn wide attention these days. And \textsc{Nettack} was proposed to generate adversarial examples with respect to graph structure and node feature to fool the GCN model~\cite{zugner2018adversarial}. Another gradient-based method called fast gradient attack~(FGA) makes full use of the gradients information to choose candidate links that need modification when performing attack~\cite{chen2018fast}. Not limited to the manipulation on links, adding fake nodes~\cite{wang2018attack} could also minimize the classification accuracy of GCN. Apart from the attacks on GCN, some unsupervised embedding methods are also concerned, e.g., Projected Gradient Descent~(PGD) based attack~\cite{sun2018data} was introduced to lower the accuracy of DeepWalk~\cite{perozzi2014deepwalk} and LINE~\cite{tang2015line} on link prediction task. As innovative as they are, these attack approaches are still limited to the algorithms of static networks.

\section{Preliminary and Problem Formulation}
\label{sec:preliminary}
In this section, we present the definition of DNLP, as well as the adversarial attacks on it.
\subsection{Dynamic Network Link Prediction}
A network structure could be represented by $\mathbf{G} = \{\mathbf{V}, \mathbf{E}\}$, where $\mathbf{V}=\{v_1, v_2, \cdots, v_n\}$ denotes the set of network nodes, and $\mathbf{E}\subseteq\mathbf{V}\times{\mathbf{V}}$ represents the set of links. A directed link from $v_i$ pointing to $v_j$ is denoted by an ordered pair of nodes $(v_i, v_j)$.
In this paper, we focus on the dynamic networks with fixed node set but temporal links.
Such a dynamic network could be modeled as a sequence of graphs  \{$\mathbf{G}_{t-N}$, $\cdots$, $\mathbf{G}_{t-1}$\}, where $\mathbf{G}_k = \{\mathbf{V}, \mathbf{E}_k\}$ represents the network's structure at the $k^{th}$ interval.
With this, the definition of DNLP goes as follows.


\begin{itemize}
    \item \emph{Given a sequence of $\textit{N}$ graphs $\mathbf{S}$ = \{$\mathbf{A}_{t-N}$, $\cdots$, $\mathbf{A}_{t-1}$\}, where $\mathbf{A}_k$ denotes the adjacent matrix of $G_k$, the task of dynamic network link prediction is to learn a mapping $\mathbf{S} \to \mathbf{A}_{t}$, from historical snapshots to future network structure.}
\end{itemize}

Specifically, DNLP algorithms capture latent spatial features and temporal patterns from historical information, i.e. previous $\textit{N}$ adjacency matrices, and then are able to infer the adjacency matrix of next snapshot. A link exists between $v_i$ and $v_j$ at time $t$ if the probability $P(\mathbf{A}_{t}(i,j))$ given by the DNLP algorithm is larger than some threshold.

\begin{figure*}[!ht]
    \centering
    \includegraphics[width=0.9\linewidth]{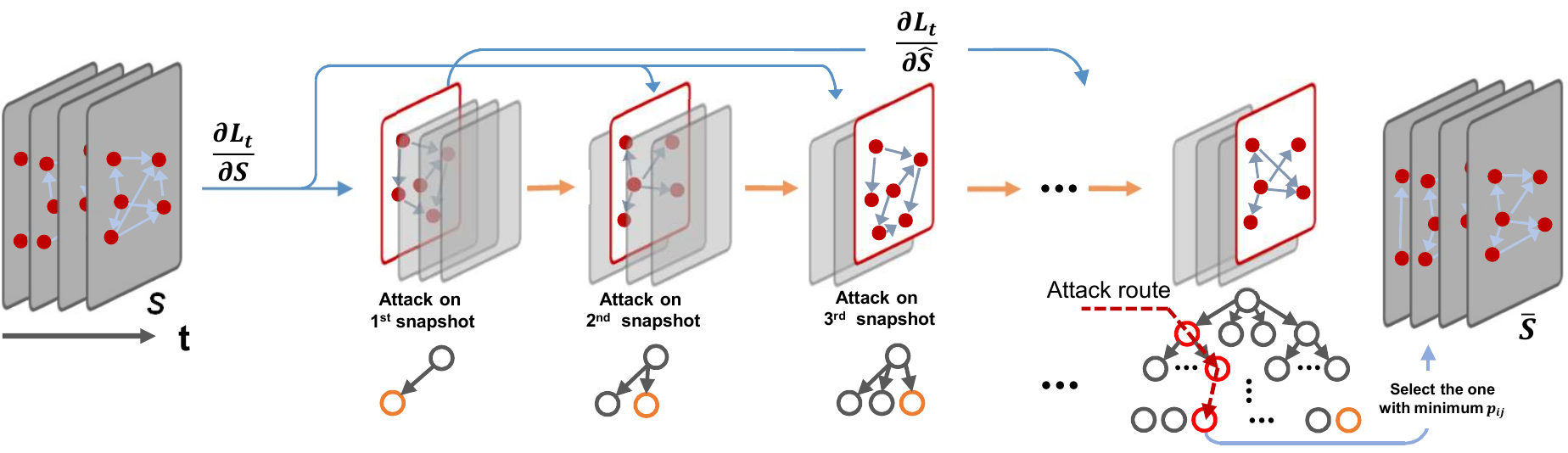}
    \caption{The framework of TGA-based methods. The nodes in the trees below represent the generation of adversarial example sets and each one in orange corresponds to the set generated in the current iteration.}
    \label{fig:framework}
\end{figure*}

\subsection{Adversarial Attack on DNLP}
The idea of the adversarial attack has been extensively explored in computer vision, which is typically achieved by adding unnoticeable perturbation to images in order to mislead classifiers. Similarly, adversarial network attack on DNLP generates adversarial examples by adding or deleting a limited number of links from the original network, so as to make DNLP algorithms fail to predict target linkages. Intuitively, the goal of the generated adversarial example is to minimize the probability of the target link predicted by the DNLP algorithms, which could be formalized as
\begin{equation}
\begin{split}
& \min P(\mathbf{A}_{t}(i,j)|\mathbf{\hat{S}}) \\
& \textrm{subject to}~~\mathbf{\hat{S}} = \mathbf{S} + \triangle{\mathbf{S}} \\
\end{split}
\end{equation}
where $\mathbf{\hat{S}}$ denotes the generated adversarial example, and $\triangle{\mathbf{S}}$ is the perturbation introduced into $\mathbf{S}$, i.e., the amount of links that need modification.

Different from the attack strategies on static network algorithms, the chosen links added to or deleted from historical snapshots are associated with temporal information. One same link on different snapshots may contribute differently in the prediction of target link, not to mention that the linkages are time-varying. Therefore, it is crucial to take \emph{time} into consideration when designing the attacks.

\section{Time-aware Gradient Attack}
\label{sec:method}
In order to generate adversarial dynamic network with optimal link modification scheme, a naive idea is to search through permutation and combination, which however is
extremely time-consuming.
Fortunately, deep learning based DNLP methods produce abundant information when making predictions, i.e., the gradients, which may assist adversarial example generation. Here, we first briefly introduce DDNE and then show how it can help to generate adversarial examples. The framework is shown in Fig.~\ref{fig:framework}. It should be noted that it doesn't matter which DNLP model is adopted here, as long as it can achieve a reasonable performance.


\subsection{The Framework of DDNE}
DDNE has a dual encoder-decoder structure. A GRU could be used as the encoder, which reads the input node sequence both forward and backward, and turns the node into lower representation. The decoder, which consists of several fully connected layers, restores the input node from the extracted features. For a node $v_i$, the encoding process is described as
\begin{equation}
\label{equation: DDNE-ENCODER}
\begin{split}
& \overrightarrow{\mathbf{h}}_{i}^{k} = \text{GRU}(\overrightarrow{\mathbf{h}}_{i}^{k-1} + \overrightarrow{\mathbf{S}}^{k}(i:,)), \\
& \overleftarrow{\mathbf{h}}_{i}^{k} = \text{GRU}(\overleftarrow{\mathbf{h}}_{i}^{k-1} + \overleftarrow{\mathbf{S}}^{k}(i:,)), \\
& \mathbf{h}_{i}^{k} = [\overrightarrow{\mathbf{h}}_{i}^{k}, \overleftarrow{\mathbf{h}}_{i}^k],~~~~k=\{t-N, \cdots, t-1\}, \\
& \mathbf{c_i} = [\mathbf{h}_{i}^{t-N}, \cdots, \mathbf{h}_{i}^{t-1}], \\
\end{split}
\end{equation}
where $\mathbf{h}_{i}^{k}$ represents the hidden state of the GRU when processing $v_i$ of the $k^{th}$ snapshot, and $\mathbf{c}_i$ is the concatenation of all $\mathbf{h}_{i}^{k}$ in time order. $\mathbf{h}_{i}^{k}$ consists of two parts, the forward one $\overrightarrow{\mathbf{h}}_{i}^{k}$ and the reversed one $\overleftarrow{\mathbf{h}}_{i}^{k}$, which are fed with opposite time sequence, $\overrightarrow{\mathbf{S}}(i,:))=\{\mathbf{A_{t-N}}(i,:), \cdots, \mathbf{A_{t-1}}(i,:)\}$ and $\overleftarrow{\mathbf{S}}(i:,))=\{\mathbf{A_{t-1}}(i,:), \cdots, \mathbf{A_{t-N}}(i,:)\}$, respectively. The decoder is composed of multilayer perceptrons, of which the complexity may vary according to the scale of datasets. The decoding process could be formulated as
\begin{equation}
\label{equation: DDNE-DECODER}
\begin{split}
& \mathbf{y}_{i}^{(1)} = \sigma_{1}(\mathbf{W}^{(1)}\mathbf{c}_i + \mathbf{b}^{(1)}), \\
& \mathbf{y}_{i}^{(m)} = \sigma_{m}(\mathbf{W}^{(m-1)}\mathbf{y}_{i}^{(m-1)} + \mathbf{b}^{(m)}), m=2,\cdots,M \\
\end{split}
\end{equation}
where $M$ represents the number of layers in the decoder, and $\sigma_{m}$ denotes the activation function applied in the $m^{th}$ decoder layer. Here, $\sigma_{m}=\text{ReLU}(\cdot)$ when $m<M$ and $\sigma_{M}=\text{sigmoid}(\cdot)$. In the training process, DDNE minimizes objective function $\mathcal{L}_{all}$ which consists of three parts: an adjusted $L_2$ loss $\mathcal{L}_{s}$ between predicted snapshot and the true one to learn the transition pattern, an adjusted $L_2$ loss $\mathcal{L}_{c}$ between the two embeddings to capture interaction proximity and a regularization term $\mathcal{L}_{reg}$ to avoid overfitting. And $\mathcal{L}_{all}$ is defined as
\begin{equation}
\label{equation:L_ALL}
\mathcal{L}_{all} = \mathcal{L}_{s} + \beta\mathcal{L}_{c} + \gamma\mathcal{L}_{reg}.
\end{equation}
Here, $\mathcal{L}_{s}$ adds an additional weight $\mathbf{Z}(i,:)$ to $L_2$ loss in order to ease the impact of sparsity, with $\{\mathbf{Z}(i,j)\}_{j=1}^{n} = 1$ if $\mathbf{S}^{t}(i,j) = 0$ and $\{\mathbf{Z}(i,j)\}_{j=1}^{n} = \alpha > 1$ otherwise. $\mathcal{L}_{s}$ is defined as
\begin{equation}
\label{equation:L_S}
\begin{split}
    \mathcal{L}_{s} &= -\sum_{i=1}^{n}\mathbf{Z}(i,:)\odot[\mathbf{A}_{t}(i,:) - \hat{\mathbf{A}}_{t}(i,:)]^2 \\
                    &= -\sum_{i=1}^{n}\sum_{j=1}^{n}\mathbf{Z}(i,j)[\mathbf{A}_{t}(i,j) - \hat{\mathbf{A}}_{t}(i,j)]^2.
\end{split}
\end{equation}
On the other hand, $\mathcal{L}_{c}$ imposes $N_{ij}$, the amount of links between $v_i$ and $v_j$ in historical snapshots, to $L_2$ loss. It addresses the influence of historical connections, and is defined as
\begin{equation}
\label{equation:L_C}
\mathcal{L}_{c} = \sum_{u, v=1}^{n}{N_{ij}\parallel\mathbf{c}_{i}-\mathbf{c}_{j}\parallel}_{2}.
\end{equation}

\subsection{Time-aware Link Gradient}
When training DDNE, we calculate $\partial{\mathcal{L}_{all}}/\partial{W}$ to update the corresponding weights $W$ through stochastic gradient descent. Similarly, in adversarial network generation, we can update $S(i,:)$  by taking $\partial{\mathcal{L}_{all}}/\partial{S(i,:)}$ with ${S(i,:)}$ being the variable. $\mathcal{L}_{all}$ integrates the information of the entire network, which makes the links that contribute the most in prediction covered among all links in $S(i,:)$. To find out the most valuable links in target link prediction, we design $\mathcal{L}_t$ to only take the target link into consideration. Its definition goes as follows.
\begin{equation}
\mathcal{L}_{t} = -[1-\mathbf{\hat{A}}_t(i, j)]^2,
\end{equation}
with $\mathbf{\hat{A}}_t(i, j)$ equal to $P(\mathbf{A}_{t}(i,j))$, the probability generated by DDNE. This can make the time-aware link gradient, $\partial\mathcal{L}_t/\partial{S(i,:)}$, more concentrated when it is applied in target link attack. The calculation of $\partial\mathcal{L}_t/\partial{\mathbf{S}(i,:)}$ follows the chain rule and the partial derivative can be obtained by calculating $\partial{f(\mathbf{S}(i,:))(i, j)}/\partial{\mathbf{S}(i,:)}$ with DDNE regarded as $f$, which is described as
\begin{equation}
\label{equation:tga}
\begin{split}
    \frac{\partial{\mathcal{L}_t}}{\partial{\mathbf{S}(i,:)}} &= 2[1 - \mathbf{\hat{A}}_t(i, j)]\frac{\partial\mathbf{\hat{A}}_t(i, j)}{\partial{\mathbf{S}(i,:)}}\\
                                                     &= 2[1 - \mathbf{\hat{A}}_t(i, j)]\frac{\partial{f(\mathbf{S}(i,:))(i, j)}}{\partial{\mathbf{S}(i,:)}}.
\end{split}
\end{equation}
Note that $\partial\mathcal{L}_t/\partial{\mathbf{S}(i,:)}$ is a tensor with the same shape of $\mathbf{S}(i,:)$, and the element $g_k(i,j)$ represents the gradient of linkage $(i,j)$ on the $k^{th}$ snapshot.

\begin{figure*}[!ht]
    \centering
    \includegraphics[width=1.\linewidth]{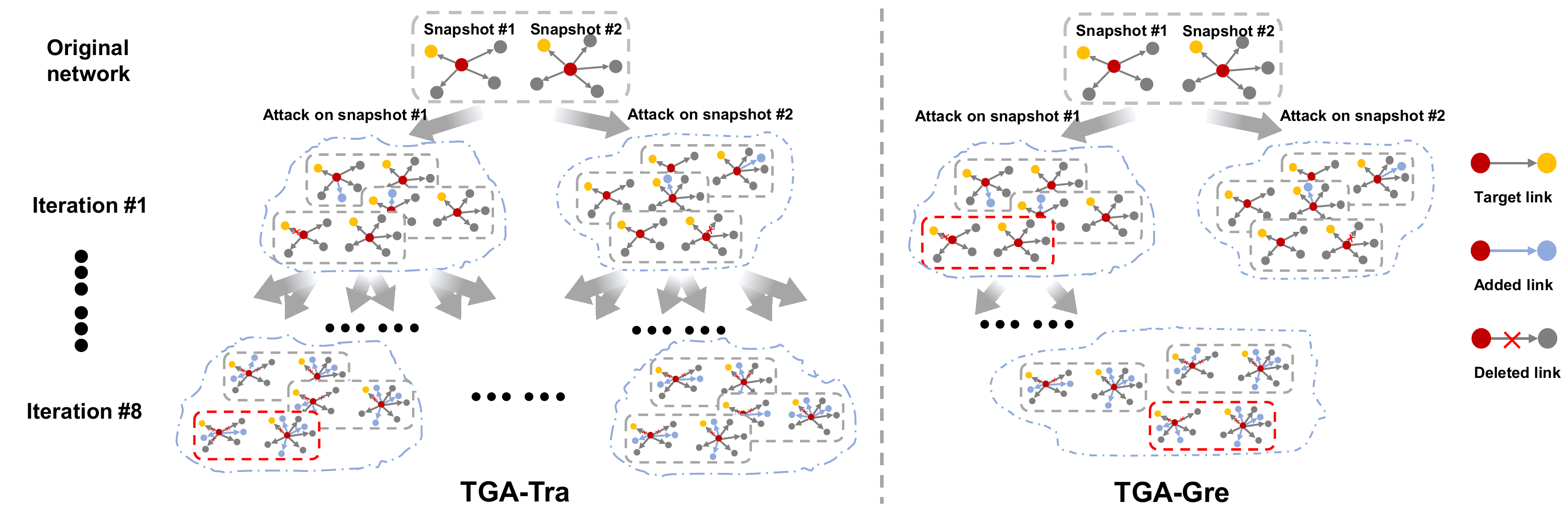}
    \caption{Illustration of TGA-Tra and TGA-Gre with $n_s=2$, $\gamma=8$ and $m=3$. The doted boxes in red represent the final adversarial examples. By comparison, TGA-Tra needs to compare a larger number of candidates, leading to higher time complexity.}
    \label{fig:tga_process}
\end{figure*}

\subsection{Traversal Search Based TGA}
Given a dynamic network $\mathbf{S}$, we aim to find $\mathbf{\triangle{S}}$ such that the target link could be prevented being predicted from DDNE at a minimum cost. Possible solutions are myriad since there are tens of thousands of nodes and multiple snapshots in $\mathbf{S}$, making it hard to find a proper $\mathbf{\triangle{S}}$. Chen et al. propose to modify links according to Eq.~(\ref{equation:modification}) to disturb GCN on the task of node classification. The modification involves both the magnitude and sign of $g(i,j)$, which decides the candidate linkages and how they should be modified, respectively.

\begin{equation}
\label{equation:modification}
\hat{\mathbf{A}}(i,j) = \mathbf{A}(i,j) + \mathit{sign}(g(i,j)).
\end{equation}
It is designed on the assumption that the link with maximum magnitude of gradient contributes the most when inference. And the sign of gradients, denoted as $\mathit{sign}(g(i,j))$, determines whether the effect is positive or negative. However, neither GCN or DDNE could learn network features and make prediction flawlessly, which might result in the deviation of gradients. It means that modifying the link with maximum magnitude of gradient does not always lead to best attack performance and $\mathit{sign}(g(i,j))$ might express the opposite meaning.

\subsubsection{Search across Snapshots} 
In TGA-based attack methods, we first group top $m$ links based on the magnitude of the gradients $\partial{\mathcal{L}_t}/\partial{\mathbf{S}(i,:)}$ and then find the optimal link to modify within the $m$ candidates. Specifically, Within one iteration, we first sort links based on $|g_{ij}|$ with respect to each snapshot, and then select top $m$ links in each snapshot. We then generate candidate adversarial examples by altering the linkage status of the selected links. Specifically, if there is a link between $v_i$ and $v_j$ at the $k^{th}$ snapshot, we then remove $e_{ij}$; In the reverse case, suppose $v_i$ and $v_j$ are not connected at the $k^{th}$ snapshot, we then can add a link between $v_i$ and $v_j$ on the corresponding snapshot. Note that we do not follow the rule defined in Eq.~(\ref{equation:modification}) but just perform attack on the basis of linkage status. When implementing the above steps, we treat the one-link modification on the target snapshot as a basic operation, called \textit{OneStepAttack}, as shown in Algorithm~\ref{algorithm: one_stap_attack}.


\begin{algorithm}[!ht]
\label{algorithm: one_stap_attack}
\SetAlgoLined
\KwIn{Original network $\mathbf{S}(i, :)$, the partial derivative $\partial\mathcal{L}_t/\partial\mathbf{S}(i,:)$, target snapshot \textit{k}, number of candidate adversarial examples $m$;}
\KwOut{A set of candidate adversarial networks $\overline{\mathbf{S}}(i,:)$}
Initialize $\overline{\mathbf{S}}(i,:)=\mathbf{S}(i,:)$\;
Initialize empty candidate adversarial examples set $\mathit{CA}$\;
Initialize $Q=\{g_k(i,0), \cdots, g_k(i,n)\}$ and sort it by the magnitudes of their gradient in ascending order.\;
\For{$i=0$ to $m$} {
Get the target link based on $Q[i]$\;
Generate a adversarial example and add it to CA;
}
return CA;
\caption{OneStepAttack}
\end{algorithm}

\subsubsection{Search between Iterations}
We iteratively add perturbations to $\mathbf{S}(i,:)$ when performing the attack. In each iteration, the perturbation $\mathbf{\triangle{S}}$ is designed according to the time-aware link gradients. We have two options for this, one is to use the gradients obtained in the last iteration (for higher efficiency), and the other is to re-calculate them based on the adversarial example generated in the last iteration (for higher effectiveness). Here, we use the second one.
We first obtain all possible adversarial examples in $\gamma$ iterations. After that, we choose the one which could achieve the minimum $p_{ij}$ as the final adversarial example. The procedure of our TGA-Tra is visualized in the left part of Fig.~\ref{fig:tga_process}. For instance, when we set the number of historical snapshots $n_s=2$, $\gamma=8$ and $m=2$, we will have $2^{15}$ possible adversarial examples as candidates, and the red doted box in Iteration 8 is the finally chosen one. Clearly we can get the attack route through backtracking. The details of TGA-Tra are presented in Algorithm~\ref{algorithm: tga_tra}.


\begin{algorithm}[!ht]
\label{algorithm: tga_tra}
\SetAlgoLined
\KwIn{A trained \textit{DDNE}, original network $\mathbf{S}{(i, :)}$, number of modifications $\gamma$, number of candidate adversarial examples $m$;}
\KwOut{adversarial example: $\hat{\mathbf{S}}(i,:)$}
Initialize candidate adversarial examples set $\mathit{CA}={\mathbf{S}(i,:)}$\;
\While{$\gamma > 0$} {
    Initialize empty set $\overline{CA}$\;
    \For{each adv\_example in CA}{
        \For{$k=1$ to $n_s$}{
            g = $\partial\mathcal{L}_t/adv\_example$\;
            $\mathbf{\hat{S}}$ = \textit{OneStepAttack}(\textit{adv\_example}, g, k, m)\;
            Add $\mathbf{\hat{S}}$ to $\overline{CA}$\;
        }
    }
    CA = $\overline{CA}$\;
    $\gamma = \gamma - 1$\;
}
\For{each adv\_example in CA}{
    $p(i,j) = DDNE(adv\_example)$\;
}
Select $\hat{\mathbf{S}}(i,:)$ as the one with minimum p(i,j) in \textit{CA}\;
return $\hat{\mathbf{S}}(i,:)$;
\caption{TGA-Tra: Attack via traversal search}
\end{algorithm}

\subsection{Greedy Search Based TGA}
\label{sec: tga_gre}
TGA-Tra could be effective since it 
compares a large number of modification schemes, 
but it is of relatively high time complexity, especially for large $N$ and $n_s$. Taking the partial derivatives of the input pairs of nodes~(\textit{op1}) and sorting them in descending order~(\textit{op2}) are the two most time-consuming steps when performing the attack. For TGA-Tra, we need to repeat the above two steps at most $m^{\gamma-1}n_{s}^{\gamma}$ times to hide the target link from being predicted, 
which is barely affordable in real cases. We thus propose another  greedy search method, namely TGA-Gre, 
as shown in the right part of Fig.~\ref{fig:tga_process}. Here, in each iteration, we select the one achieving the minimum $p_{ij}$ as the most effective adversarial example, which is further considered as the input of next iteration. The details of TGA-Gre are presented in Algorithm~\ref{algorithm: tga_gre}.

\begin{algorithm}[!t]
\label{algorithm: tga_gre}
\SetAlgoLined
\KwIn{A trained \textit{DDNE}, original network $\mathbf{S}{(i, :)}$, target link (i,j), number of modifications $\gamma$;}
\KwOut{adversarial example: $\hat{\mathbf{S}}(i,:)$}
Initialize $\mathbf{\hat{S}}(i,:) = \mathbf{S}(i,:)$\;
\While{$\gamma > 0$} {
    Initialize empty candidate adversarial example set \textit{CA}\;
    \For{$k=1$ to $n_s$}{
        g = $\partial\mathcal{L}_t/\mathbf{\hat{S}}(i,:)$\;
        $\mathbf{\hat{S}}$=\textit{OneStepAttack}($\mathbf{\hat{S}}(i,:)$, g, k)\;
        Add $\mathbf{\hat{S}}_{temp}(i,:)$ to CA\;
        $p(i,j) = DDNE(\mathbf{\hat{S}})$\;
    }
    Select $\hat{\mathbf{S}}(i,:)$ as the one with minimum p(i,j) in \textit{CA}\;
    $\gamma$ = $\gamma - 1$\\
}
return $\hat{\mathbf{S}}(i,:)$\\
\caption{TGA-Gre: Attack via greedy search}
\end{algorithm}

For a target link $(i, j)$, TGA-Gre assumes that the lowest $p_{ij}$ in each iteration could lead to the best attack result. It avoids massive comparisons between iterations and thus can significantly accelerate the whole process. Similar to the procedure of TGA-Tra, TGA-Gre also compares $g_t$ across all snapshots during each iteration. The major difference is that TGA-Gre elects a local optimal link in each iteration. 
It is clear that we only need to repeat \textit{op1} and \textit{op2} at most $\gamma n_{s}$ times, which is independent of $m$ and makes the attack much more efficient. 

\section{Experiments}
\label{sec:experiment}

\subsection{Datasets}
We perform experiments on four real-world dynamic networks, 
with their basic statistics listed in Table~\ref{dataset}.
\begin{itemize}
  \item \textsc{\textbf{radoslaw}}~\cite{konect:2017:radoslaw}: It is an internal email network between employees in a mid-sized manufacturing company. We focus on the nodes appeared in 2010-01 and construct network using their interactions from 2010-02-01 to 2010-09-30.
  \item \textbf{\textsc{lkml}}~\cite{konect:2016:lkml-reply}: It is also an email network from the linux kernel mailing list. We focus on the users who appeared on the mailing list between 2007-01-01 and 2007-04-01 and slice the data during the next 12 months at an interval of 3 months.
  \item \textbf{\textsc{dnc}~\cite{konect:2016:dnc-temporalGraph}} It is a directed network of emails in the 2016 Democratic National Committee email leak. We construct the dynamic network using the emails between 2016-04-29 and 2016-05-23 with the nodes appeared during 2016-04-19 and 2016-05-02.
  \item \textsc{\textbf{enron}}~\cite{graphrepository2013}: It is an email network covering decades of the email communication in Enron. We use the records between 2000-04-01 and 2001-04-01 experiments and slice them into 4 snapshots in 3-month increments. Also, we only focus on part of the email address that at least sent an email from 2000-01-01 to 2000-04-01.
\end{itemize}
As described, all the datasets are divided into at least 7 snapshots with different intervals. The first snapshot provides the nodes we need to focus and the rest are used for training and inference. All the datasets are available online $\footnote{http://konect.uni-koblenz.de/}$.

\begin{table}
\centering
\caption{The basic statistics of the three datasets}
\label{dataset}
\renewcommand{\arraystretch}{1.1}
\begin{tabular}{cccccc}
\hline \hline
Dataset                 & $|V|$   & $|{E}_{T}|$  & $\bar{d}$ & $d_{max}$  & \begin{tabular}[c]{@{}c@{}}Timespan\\ (days)\end{tabular}\\
\hline
\textsc{radoslaw}       & 151     & 72.2K        & 27.7      & 240        & 242      \\ 
\textsc{dnc}            & 1,891   & 46.7K        & 1.24      & 198        & 731      \\ 
\textsc{lkml}           & 2,210   & 201.7K       & 7.9       & 718        & 731      \\ 
\textsc{enron}          & 2,628   & 308.8K       & 2.23      & 228        & 365      \\ 
\hline \hline
\end{tabular}
\end{table}

\begin{table*}[!t]
\renewcommand{\arraystretch}{1.1}
\centering
\caption{Attack performance in terms of ASR and AML}
\begin{tabular}{l|ccc|cc|ccc|cc}
\hline \hline
\multirow{2}{*}{Dataset} & \multicolumn{5}{c|}{ASR}                & \multicolumn{5}{c}{AML}                \\ \cline{2-11}
                         & RA         & CNA       & FGA        & TGA-Gre    & TGA-Tra    & RA         & CNA          & FGA          & TGA-Gre      & TGA-Tra      \\ \hline
\multicolumn{11}{c}{Attack on top-100 links with the highest existence probability}                                                                               \\ \hline
\textsc{radoslaw}        & 0.01       & 0.32      & 0.68       & 1.00       & 1.00       & 9.95       & 8.76         & 5.06         & 2.65         & 2.32         \\
\textsc{dnc}             & 0.00       & 0.40      & 0.54       & 0.93       & 0.94       & 10.00      & 8.93         & 8.54         & 5.37         & 5.06         \\
\textsc{lkml}            & 0.00       & 0.19      & 0.47       & 0.73       & 0.77       & 10.00      & 8.02         & 8.31         & 5.71         & 5.29         \\
\textsc{enron}           & 0.00       & 0.13      & 0.30       & 0.83       & 0.85       & 10.00      & 8.94         & 8.66         & 4.15         & 3.82         \\ \hline
\multicolumn{11}{c}{Attack on top-100 links with the highest degree centrality}                                                                                   \\ \hline
\textsc{radoslaw}        & 0.18       & 0.60      & 0.99       & 0.99       & 1.00       & 8.78       & 6.72         & 2.70         & 1.69         & 1.67         \\
\textsc{dnc}             & 0.06       & 0.47      & 0.83       & 0.90       & 0.92       & 9.82       & 7.82         & 5.91         & 4.50         & 4.17         \\
\textsc{lkml}            & 0.00       & 0.21      & 0.30       & 0.38       & 0.41       & 10.00      & 9.13         & 8.98         & 7.75         & 7.19         \\
\textsc{enron}           & 0.00       & 0.59      & 0.86       & 0.86       & 0.87       & 10.00      & 6.98         & 5.97         & 5.97         & 5.69         \\ \hline
\multicolumn{11}{c}{Attack on top-100 links with the highest edge betweenness centrality}                                                                                                                      \\ \hline
\textsc{radoslaw}        & 0.30       & 0.62      & 1.00       & 1.00       & 1.00       & 8.44       & 7.23         & 1.70         & 1.48         & 1.48         \\
\textsc{dnc}             & 0.01       & 0.71      & 0.89       & 0.95       & 0.95       & 9.94       & 6.21         & 5.45         & 4.32         & 4.15         \\
\textsc{lkml}            & 0.04       & 0.48      & 0.75       & 0.77       & 0.79       & 9.74       & 6.83         & 4.07         & 3.82         & 3.82         \\
\textsc{enron}           & 0.07       & 0.57      & 0.92       & 0.98       & 0.98       & 9.67       & 5.49         & 3.15         & 2.19         & 2.12         \\
\hline \hline
\end{tabular}
\label{table:results}
\end{table*}

\subsection{Baseline Methods}
As the first work to study adversarial attack on DNLP algorithms, we design three baseline attacks as follows, to compare with TGA-Gre and TGA-Tra.
\begin{itemize}
\item \textbf{Random Attack~(RA)}: RA randomly modifies $\gamma$ linkages in all snapshots. In practice, we add $b$ new connections to the target node and remove $\gamma-b$ links between the target node and its neighbors. Here, we use RA to see the robustness of DNLP algorithms under random noises.
\item \textbf{Common-Neighbor-based Attack~(CNA)}: CNA adds $b$ links between node pairs with less common neighbors and remove $\gamma-b$ links between those with more common neighbors. We adopt CNA as a baseline since \emph{common neighbor} is the basis of many similarity metrics between pairwise nodes used for link prediction.
\item \textbf{Fast Gradient Attack~(FGA)}~\cite{chen2018fast}: FGA recursively modify the linkage with maximum absolute gradient obtained by $\partial{L}_{t}/\partial{\mathbf{S}(i,:)}$, until the attack succeeds, or the number of modifications reaches $\gamma$. We use FGA as a baseline to address the importance of utilizing temporal information in attacking DNLP algorithms.
\end{itemize}

We set $\gamma=10$ in all experiments and $b=5$ for RA and CNA. 
For TGA-based methods, we set $m$ to 5. We first train the models with original data, and then feed adversarial examples generated by DDNE to each model, to validate the effectiveness of the attacks.

\subsection{Evaluation Metrics}
We choose \textit{attack success rate}~($\mathcal{ASR}$) and \textit{average attack modification links}~($\mathcal{AML}$) as attack effectiveness criterion. 

\begin{itemize}
  \item \textbf{$\mathcal{ASR}$}: The ratio of the amount of links that are successfully hidden 
  to the total number of target links that can be correctly predicted in the target snapshot. 
  \item \textbf{$\mathcal{AML}$}: The average amount of perturbation to prevent each target link from being predicted. If it needs to modify at least  $q_{i}$ links to hide link $i$, then $\mathcal{AML}$ is defined as
      \begin{equation}
        \label{equation:aml}
        \mathcal{AML} = {\frac{1} {m}}{\sum_{i=0}^{m-1} {q_i}},
      \end{equation}
      where $m$ represents the number of target links. Note that $q_{i}\leq\gamma$ and the equality holds when the attack fails.
\end{itemize}
We use $\mathcal{ASR}$ to evaluate attack methods in the first place which reflects the possibility to successfully perform attack and then compare their $\mathcal{AML}$ when their $\mathcal{ASR}$ are close.

\subsection{Results}
Firstly, we use the generated adversarial examples to fool the DDNE model to prevent target links from being predicted.
We set $\gamma=10$ to ensure the disguise of modification, which also leads the maximum of $\mathcal{AML}$ equal to 10. The results are presented in Table~\ref{table:results}. The two TGA methods outperform FGA in terms of both $\mathcal{ASR}$ and $\mathcal{AML}$, while FGA is better than CNA and RA. The results suggest that: 1) the gradients of DDNE is critical to attack different DNLP methods; 2) utilizing temporal information can indeed significantly improve the attack effect. Moreover, we study the attack performance on 3 different types of links: the links that are most likely to exist according to DDNE, the links with highest degree, in terms of the sum of terminal-node degrees, and the links with highest edge betweenness centrality. As excepted, the links with maximum existing probability are most difficult to attack in most situations. The other 2 types of links which have physical meaning in real scenarios are easier to be hidden from detection, reflecting the practicability of TGA-based methods.

As expected, TGA-Tra behaves better than TGA-Gre, but the latter is much more efficient and thus more practical in real-world applications. Almost all the methods behave much better on \textsc{fb-wosn} than the other two networks, indicating that sparser networks are relatively easier to be disturbed by adversarial attacks, that is, the DNLP algorithms on sparser networks are less robust.
The gap of the performance between TGA-Tra and TGA-Gre overturns the hypothesis that the greatest drop of $\mathcal{L}_t$ in each iteration does not lead to the best attack performance sometimes. This enlightens us to further explore specific meanings behind $g_t$. Fig.~\ref{fig:tga_process} visualizes the attack schemes of TGA-Tra and TGA-Gre performed on $E(10,4)$ of \textsc{radoslaw} on $\#4$ snapshot. We find that the performance of TGA-Tra and TGA-Gre are very similar in each iteration, but their routes seem totally different. By investigating these adversarial examples, we have the following two observations:
\begin{itemize}
    \item First, TGA-Tra is more likely to modify the links on earlier historical snapshots, while TGA-Gre tends to change the links on the most recent ones;
    \item Second, TGA-Tra prefers to add rather than delete links, while TGA-Gre has the opposite tendency.
\end{itemize}
Such observations indicate that TGA-Tra could be more concealed than TGA-Gre, since people tend to pay more attention to recent events, e.g., link change in recent snapshots. On the other hand, TGA-Gre may be preferred if we want to get some short-term attack effect. Besides, TGA-Tra seems to have lower social cost, since adding links are always easier than deleting in our social circle. Since TGA-Gre has similar performance, while is much more efficient, compared with  TGA-Tra, we will mainly focus on TGA-Gre in the rest of this paper.

\begin{figure*}
    \centering
    \subfigure[Attack process of TGA-Tra]{
    \begin{minipage}[b]{1.\textwidth}
    \includegraphics[width=1.\linewidth]{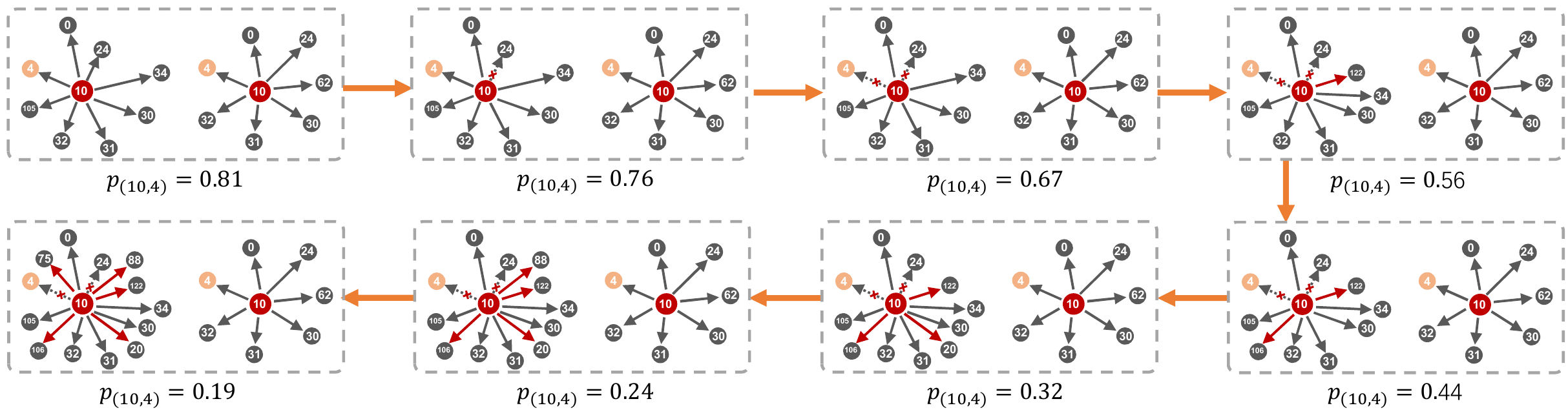}
    \end{minipage}
    }

    \subfigure[Attack process of TGA-Gre]{
    \begin{minipage}[b]{1.\textwidth}
    \includegraphics[width=1.\linewidth]{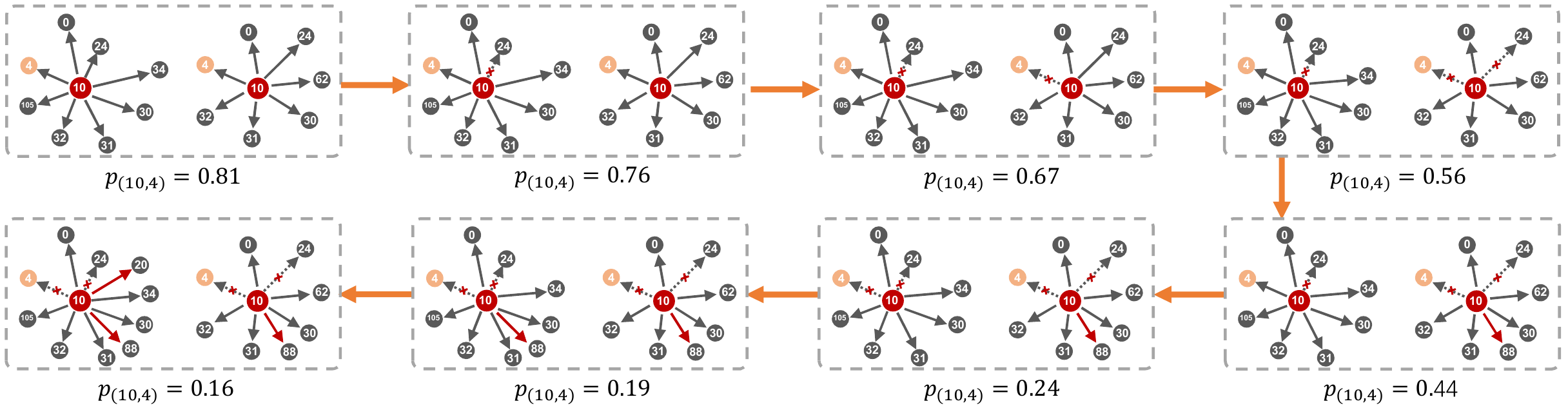}
    \end{minipage}
    }
\caption{Attack process of TGA-Tra and TGA-Gre on $E(10,4)$ of \textsc{radoslaw} on \#4 snapshot.}
\label{my_label}
\end{figure*}


\subsubsection{Attack on Long-term Prediction}
Besides focusing on the next immediate snapshot, researchers always look into the performance of DNLP algorithms on long-term prediction. Similarly, we would also like to investigate whether the TGA-based methods are effective for hiding remote future links. We first make predictions for the $\#4$, $\#5$ and $\#6$ snapshot with $n_s = 2$ and then compare the attack performance on different snapshots, as shown in Table~\ref{table:long_term_attack}. We can see that, generally, the performance of TGA-Gre are close in spite that the target snapshots vary. Additionally, there are no significant variation tendency of the attack performance. The results show that TGA-based attack methods are applicable for long-term prediction attack.

\begin{table}[!ht]
\renewcommand{\arraystretch}{1.1}
\centering
\caption{The performance of TGA-Gre with $n_s=10$}
\begin{tabular}{c|ccc|ccc}
\hline \hline
\multirow{2}{*}{Dataset}   & \multicolumn{3}{c|}{ASR}       & \multicolumn{3}{c}{AML}     \\ \cline{2-7}
                           & \#3      & \#4      & \#5     & \#3      & \#4      & \#5   \\ \hline
\multicolumn{7}{c}{Attack on top-100 links with the highest existence probability}       \\ \hline
\textsc{radoslaw}          & 1.00     & 1.00     & 0.99    & 2.65     & 2.77     & 2.99  \\
\textsc{dnc}               & 0.93     & 0.96     & 0.96    & 5.37     & 5.59     & 5.24  \\
\textsc{lkml}              & 0.73     & 0.85     & 0.83    & 5.71     & 4.42     & 4.84  \\
\textsc{enron}             & 0.83     & 0.78     & 0.83    & 4.15     & 5.28     & 3.68  \\ \hline
\multicolumn{7}{c}{Attack on top-100 links with the highest degree centrality}           \\ \hline
\textsc{radoslaw}          & 0.99     & 0.99     & 1.00    & 1.69     & 2.32     & 2.13  \\
\textsc{dnc}               & 0.90     & 0.96     & 0.97    & 4.50     & 5.14     & 3.97  \\
\textsc{lkml}              & 0.38     & 0.41     & 0.51    & 7.75     & 7.38     & 6.96  \\
\textsc{enron}             & 0.86     & 0.86     & 0.87    & 5.97     & 8.25     & 6.25  \\ \hline
\multicolumn{7}{c}{Attack on top-100 links with the highest edge betweenness centrality} \\ \hline
\textsc{radoslaw}          & 1.00     & 1.00     & 1.00    & 1.48     & 1.53     & 1.56  \\
\textsc{dnc}               & 0.95     & 1.00     & 0.99    & 4.32     & 3.38     & 3.78  \\
\textsc{lkml}              & 0.77     & 0.80     & 0.74    & 3.82     & 3.37     & 4.03  \\
\textsc{enron}             & 0.98     & 0.99     & 0.99    & 2.19     & 2.30     & 2.25  \\ \hline \hline
\end{tabular}
\label{table:long_term_attack}
\end{table}

\subsubsection{Long-history Attack}
The number of historical snapshots, $n_s$, is one of the most significant parameters that affect the performance of DNLP algorithms. Typically, larger $n_s$ means more historical information can be used in prediction, and thus may improve the performance of DNLP algorithms. Higher accuracy will equip the model with more precise gradients. Here, we are interested in whether the increase of $n_s$ will get in the way of adversarial attacks or behave the opposite.

\begin{figure}[!t]
\includegraphics[width=1.\linewidth]{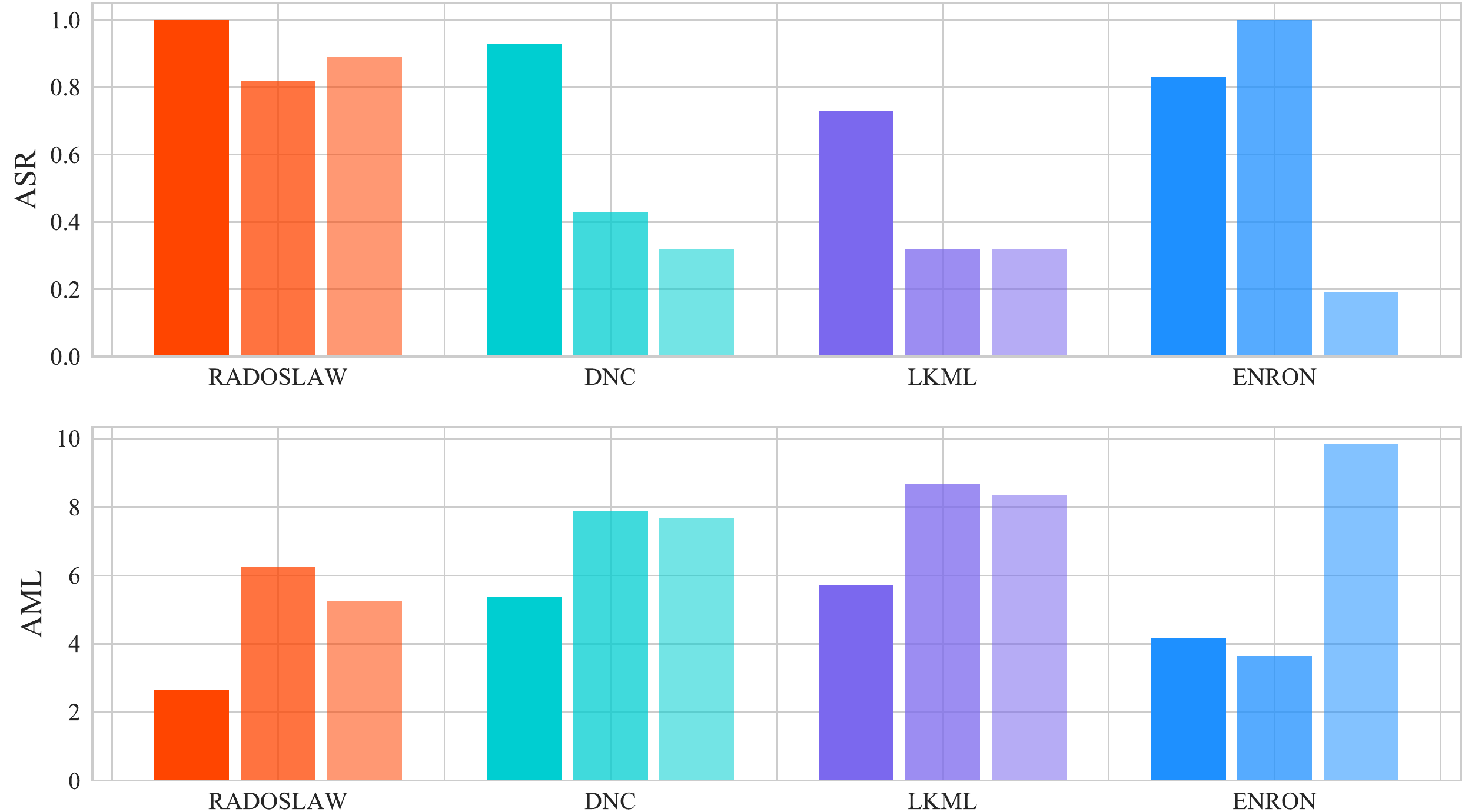}
\caption{The performance of TGA-Gre when $n_s$ changes from 2 to 4~(From left to right).}
\label{fig:ns_influence}
\end{figure}

We compare the attack performance of TGA-Gre on the 4 datasets with respect to different $n_s$. In particular, we first apply DDNE with the input sequence varying from 2 to 4, and then generate adversarial examples with $\gamma=10$.
 The results are shown in Fig.~\ref{fig:ns_influence}, where we can see that, indeed, the attack performance of TGA-Gre decrease more or less as $n_s$ increases, indicating that larger $n_s$ makes DDNE more robust against adversarial attacks. One possible reason is that the increase of $n_s$ expands the solution space and $\gamma=10$ is not enough for successfully making attacks. Actually, the performance of TGA-Gre will increase if we set $\gamma=15$. Different from the performance on the other datasets, the attack performance increase slightly when applying TGA-Gre to \textsc{enron} $n_s=3$. We argue that it is because $n_s=3$ is better for DDNE to do inference when it comes to \textsc{enron} and thus the gradients is more informative so that TGA-Gre becomes more effective. However, when $n_s$ grows to 4, the limitation of $\gamma$ shows up.

\subsubsection{Adding-link Attack}
\label{sec:adding-links}
In social networks, it is considered that deleting links is of higher social cost than adding. Moreover, temporal networks may also have multiple interactions between a pair of nodes. Therefore, deleting one link on a snapshot always removes all the corresponding interactions in the given interval. And some links may be too important to be deleted in real scenarios. Due to this gap between the cost of deleting and adding links, we would like to investigate how the attack performance of TGA-Gre will be influenced if we just add, rather than rewire, links to the original networks. The results are presented in Table~\ref{table:tga_gre_add}, where we find that the performance of TGA-Gre slightly decrease when we perform the attack only by adding links. Such results indicate that, in certain cases, we can perform the \emph{cheap} attack on DNLP by only adding a small number of links, at the cost of losing a little bit attack performance.

Also, we are surprised that the performance of TGA-Gre with only adding links on \textsc{lkml} slight increase. It shows that TGA-Gre actually find the best modification schemes on different networks and rewiring links might be the optimal choice in all cases.

\begin{table}[!t]
\renewcommand{\arraystretch}{1.1}
\centering
\caption{The performance of TGA-Gre with only adding links}
\begin{tabular}{p{1.0cm}|p{1.3cm}<{\centering}|p{1.3cm}<{\centering}|p{1.3cm}<{\centering}|p{1.3cm}<{\centering}}
\hline \hline
Dataset            & \textsc{radoslaw}  & \textsc{dnc}  & \textsc{lkml} & \textsc{enron} \\ \hline
\multicolumn{5}{c}{Attack on top-100 links with the highest existence probability}       \\ \hline
$\mathcal{ASR}$    & 1.00               & 0.88          & 0.76          & 0.69           \\
\textit{GAIN}      & -0.01              & -0.05         & 0.03          & -0.14          \\ \hline
$\mathcal{AML}$    & 4.99               & 5.86          & 5.66          & 7.26           \\
\textit{GAIN}      & 2.34               & 0.49          & -0.05         & 3.11           \\ \hline
\multicolumn{5}{c}{Attack on top-100 links with the highest degree centrality}           \\ \hline
$\mathcal{ASR}$    & 1.00               & 0.77          & 0.42          & 0.20           \\
\textit{GAIN}      & 0.01               & -0.13         & 0.04          & -0.66          \\ \hline
$\mathcal{AML}$    & 2.70               & 5.14          & 7.70          & 9.55           \\
\textit{GAIN}      & 1.01               & 0.64          & -0.05         & 3.58           \\ \hline
\multicolumn{5}{c}{Attack on top-100 links with the highest edge betweenness centrality} \\ \hline
$\mathcal{ASR}$    & 1.00               & 0.87          & 0.80          & 0.89           \\
\textit{GAIN}      & 0.00               & -0.08         & 0.03          & -0.09          \\ \hline
$\mathcal{AML}$    & 1.81               & 4.56          & 3.77          & 3.24           \\
\textit{GAIN}      & 0.33               & 0.24          & -0.05         & 1.05           \\ \hline \hline
\end{tabular}
\label{table:tga_gre_add}
\end{table}

\subsection{Runtime Comparison}
\begin{figure}[!ht]
    \centering
    \includegraphics[width=0.9\linewidth]{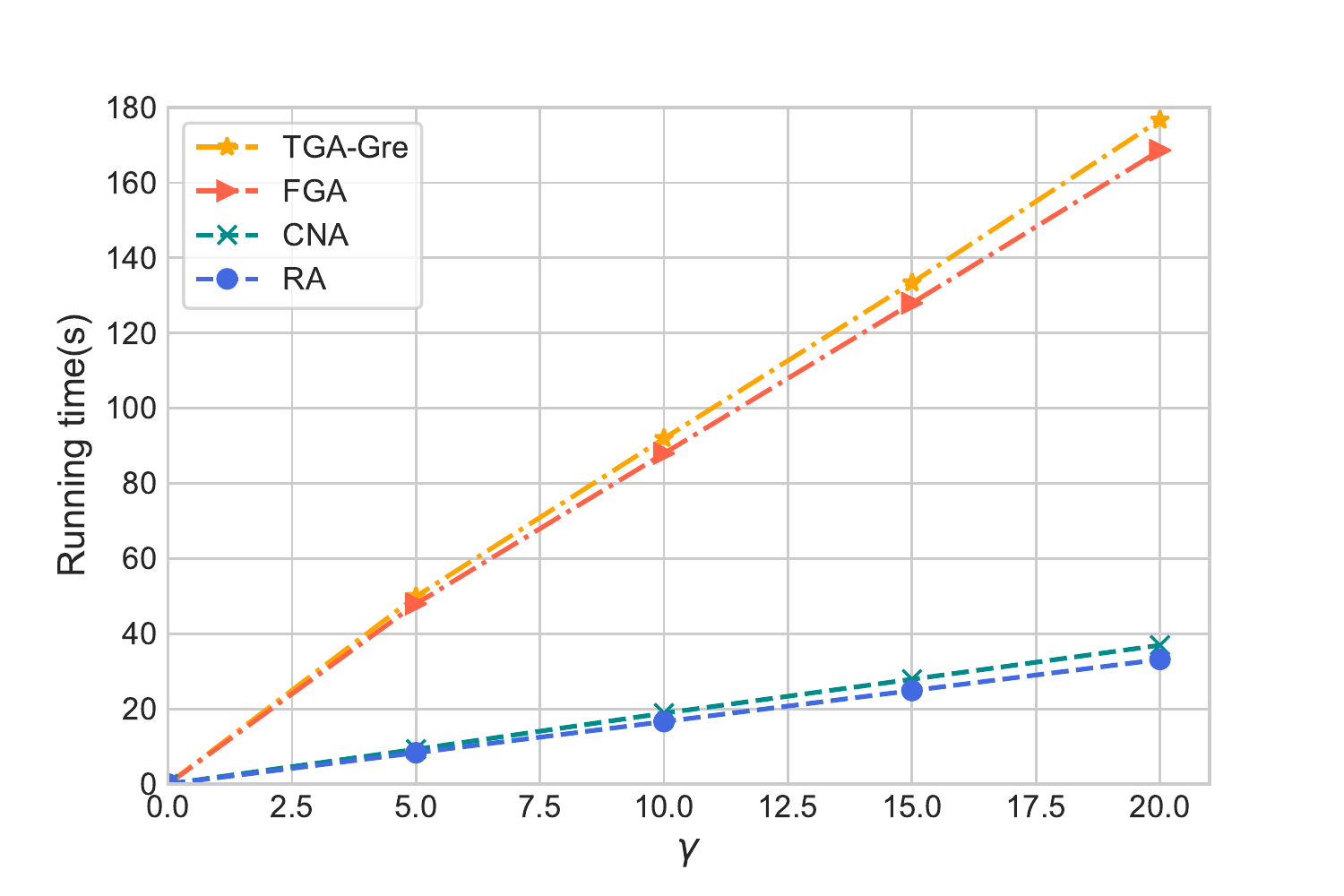}
    \caption{Runtime of different attack methods performed on the top-100 links with the highest existence probability of \textsc{dnc}.}
    \label{fig:run_time}
\end{figure}

As mentioned in Sec.~\ref{sec:method}, we propose TGA-Gre to reduce the computational complexity to make the attack more practical. To highlight the efficiency of TGA-Gre, we compare the runtime of different attack methods performed on the top-100 links with the highest existence probability of \textsc{dnc} with respect to increasing $\gamma$. In Fig.~\ref{fig:run_time}, we can find that the runtime of TGA-Gre and FGA are comparable while TGA-Gre is more effective in practice. As for RA and CNA, they have poor attack performance despite the short runtime. In fact, if the attack is performed with early stop, that is the attack process stops when it succeeds, TGA-Gre will have shorter runtime than FGA.

\subsection{Case Study: Ethereum Transaction Network}
Above experiments conducted on four benchmark datasets have shown the effectiveness of TGA-Tra and TGA-Gre on DDNE. In this section, we apply TGA-Gre on the Ethereum transaction network to hide specific transactions from detection. Ethereum is a public blockchain-based platform with the support of smart contract. With around 300 billion USD market capitalization and over 500 billion USD monthly transaction volume, it becomes the largest virtual currency trading platform second to Bitcoin. A bunch of researchers have mined the valuable data with the help of graph analysis~\cite{chen2018understanding, wu2019t,lin2020modeling} among which the analysis of temporal links, i.e. the transactions, is one of the research emphases~\cite{wu2019t}.

\begin{figure}[!t]
\centering
\includegraphics[width=1.\linewidth]{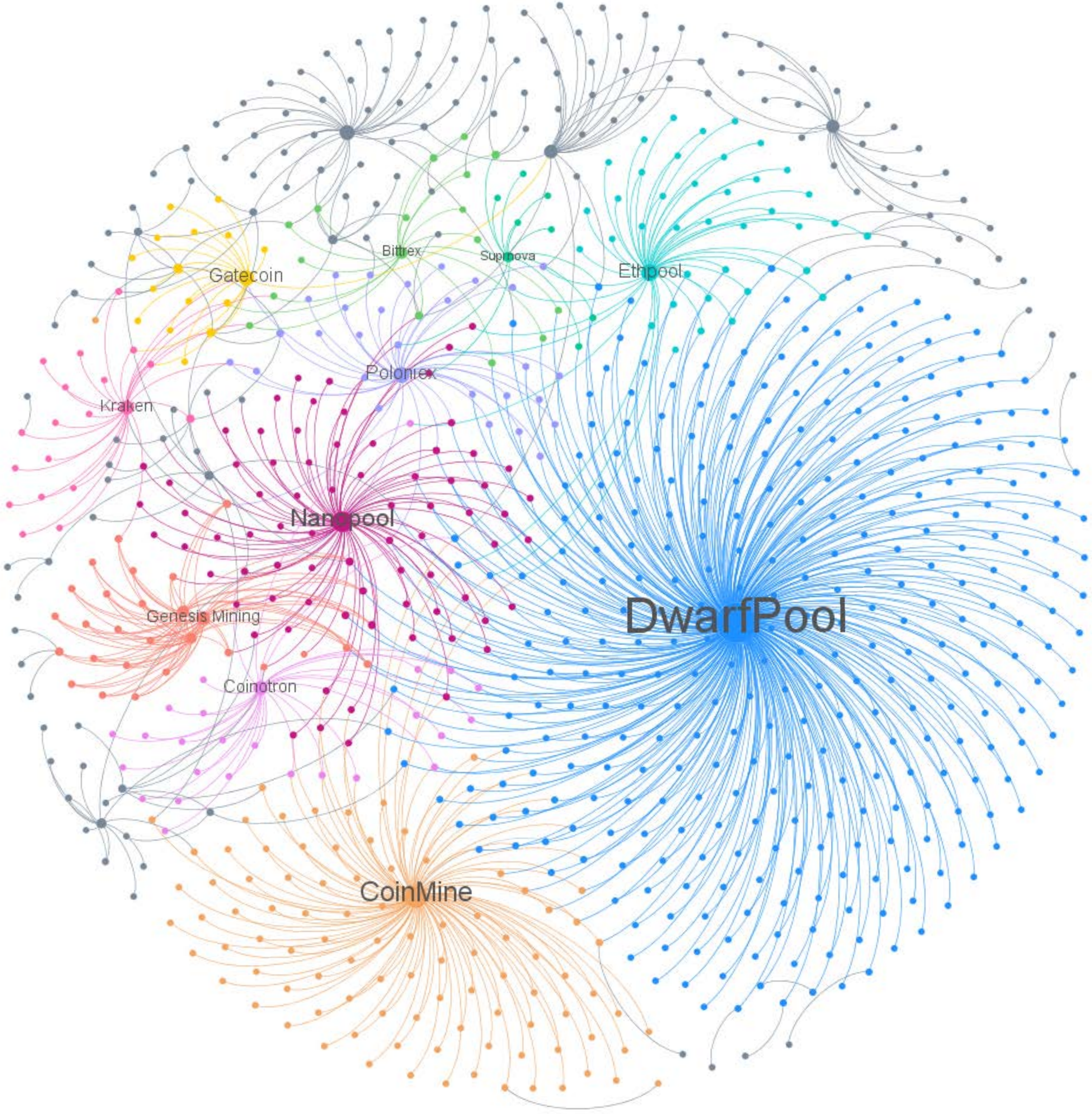}
\caption{Network structure of \#3 snapshot of Ethereum network.}
\label{fig:ethereum_network}
\end{figure}

We use the data provided by XBlock$\footnote{http://xblock.pro}$ and extract the transaction records between 2016-02 and 2016-06. The data are sliced into 5 snapshots at the interval of 1 month and
modeled as a dynamic network with 2866 nodes which represent the transaction addresses. We focus on two types of accounts: normal accounts and those belong to Ethereum pool which are identified according to the records on Etherscan$\footnote{https://etherscan.io/}$. Fig.~\ref{fig:ethereum_network} visualizes \#3 snapshot of the Ethereum transaction network from which we can find clusters centered at those Ethereum pool accounts. Applying TGA-Gre to the network, we achieve the attack performance of $\mathcal{ASR}=0.81$ and $\mathcal{AML}=5.61$ on a well-trained DDNE.

As we can observe in Fig.~\ref{fig:ethereum_network}, most links in the network are those between normal addresses and Ethereum-pool-belonged addresses which are the way normal users making profits from the pools. In real scenarios, we do not pay attentions to these links. Instead, the links between normal addresses are noteworthy. Suppose a user want to make a vital transaction reconditely in the near future, he or she could make transfers to certain addresses with the guidance of TGA. Then the transaction will not be discovered from current time scope even with the help of DDNE.

\section{Conclusion}
\label{sec:conclusion}
In this paper, we present the first work of adversarial attack on DNLP, and propose the time-aware gradient, as well as two TGA methods, namely TGA-Tra and TGA-Gre, to realize the attack. Comprehensive experiments have been carried out on four email networks and also the Ethereum transaction network.  The results show that our TGA methods behave better than the other baselines, achieving the state-of-the-art attack performance on DNLP. Besides, we investigate into the performance of TGA-Gre in several typical occasions in DNLP, including long-term prediction and long-history prediction.  Interestingly, we also find that long-term prediction seems to be more vulnerable to adversarial attacks, while using longer historical information can enhance the robustness of DNLP algorithms. The results of adding-link attack also prove the practicability of TGA-Gre.

Currently, TGA methods rely on the gradients and it requires researchers know every detail of the target DNLP model which limits its application scenarios. In the future, we would like to study the problem of \textit{black-box attack} on DNLP and further propose better strategies to improve their attack performance; on the other hand, we will also seek for methods to defend against such adversarial attacks, to achieve more robust DNLP algorithms.


%




\ifCLASSOPTIONcaptionsoff
  \newpage
\fi


\bibliographystyle{IEEEtran}
\bibliography{ref}

\begin{thebibliography}{10}
\providecommand{\url}[1]{#1}
\csname url@samestyle\endcsname
\providecommand{\newblock}{\relax}
\providecommand{\bibinfo}[2]{#2}
\providecommand{\BIBentrySTDinterwordspacing}{\spaceskip=0pt\relax}
\providecommand{\BIBentryALTinterwordstretchfactor}{4}
\providecommand{\BIBentryALTinterwordspacing}{\spaceskip=\fontdimen2\font plus
\BIBentryALTinterwordstretchfactor\fontdimen3\font minus
  \fontdimen4\font\relax}
\providecommand{\BIBforeignlanguage}[2]{{%
\expandafter\ifx\csname l@#1\endcsname\relax
\typeout{** WARNING: IEEEtran.bst: No hyphenation pattern has been}%
\typeout{** loaded for the language `#1'. Using the pattern for}%
\typeout{** the default language instead.}%
\else
\language=\csname l@#1\endcsname
\fi
#2}}
\providecommand{\BIBdecl}{\relax}
\BIBdecl

\bibitem{gong2019exact}
Y.~Gong, Y.~Zhu, L.~Duan, Q.~Liu, Z.~Guan, F.~Sun, W.~Ou, and K.~Q. Zhu,
  ``Exact-k recommendation via maximal clique optimization,'' \emph{arXiv
  preprint arXiv:1905.07089}, 2019.

\bibitem{zhang2016correlation}
S.~Zhang, J.~Ai, and X.~Li, ``Correlation between the distribution of software
  bugs and network motifs,'' in \emph{2016 IEEE International Conference on
  Software Quality, Reliability and Security (QRS)}.\hskip 1em plus 0.5em minus
  0.4em\relax IEEE, 2016, pp. 202--213.

\bibitem{fionda2018community}
V.~Fionda and G.~Pirro, ``Community deception or: How to stop fearing community
  detection algorithms,'' \emph{IEEE Transactions on Knowledge and Data
  Engineering}, vol.~30, no.~4, pp. 660--673, 2018.

\bibitem{gunecs2016link}
{\.I}.~G{\"u}ne{\c{s}}, {\c{S}}.~G{\"u}nd{\"u}z-{\"O}{\u{g}}{\"u}d{\"u}c{\"u},
  and Z.~{\c{C}}ataltepe, ``Link prediction using time series of
  neighborhood-based node similarity scores,'' \emph{Data Mining and Knowledge
  Discovery}, vol.~30, no.~1, pp. 147--180, 2016.

\bibitem{ozcan2015multivariate}
A.~{\"O}zcan and {\c{S}}.~G. {\"O}{\u{g}}{\"u}d{\"u}c{\"u}, ``Multivariate
  temporal link prediction in evolving social networks,'' in \emph{2015
  IEEE/ACIS 14th International Conference on Computer and Information Science
  (ICIS)}.\hskip 1em plus 0.5em minus 0.4em\relax IEEE, 2015, pp. 185--190.

\bibitem{fu2018link}
C.~Fu, M.~Zhao, L.~Fan, X.~Chen, J.~Chen, Z.~Wu, Y.~Xia, and Q.~Xuan, ``Link
  weight prediction using supervised learning methods and its application to
  yelp layered network,'' \emph{IEEE Transactions on Knowledge and Data
  Engineering}, vol.~30, no.~8, pp. 1507--1518, 2018.

\bibitem{xuan2018modern}
Q.~Xuan, M.~Zhou, Z.~Zhang, C.~Fu, Y.~Xiang, Z.~Wu, and V.~Filkov, ``Modern
  food foraging patterns: Geography and cuisine choices of restaurant patrons
  on yelp,'' \emph{IEEE Transactions on Computational Social Systems}, vol.~5,
  no.~2, pp. 508--517, 2018.

\bibitem{li2014deep}
X.~Li, N.~Du, H.~Li, K.~Li, J.~Gao, and A.~Zhang, ``A deep learning approach to
  link prediction in dynamic networks,'' in \emph{Proceedings of the 2014 SIAM
  International Conference on Data Mining}.\hskip 1em plus 0.5em minus
  0.4em\relax SIAM, 2014, pp. 289--297.

\bibitem{li2018deep}
T.~Li, J.~Zhang, S.~Y. Philip, Y.~Zhang, and Y.~Yan, ``Deep dynamic network
  embedding for link prediction,'' \emph{IEEE Access}, vol.~6, pp.
  29\,219--29\,230, 2018.

\bibitem{chen2019lstm}
J.~Chen, J.~Zhang, X.~Xu, C.~Fu, D.~Zhang, Q.~Zhang, and Q.~Xuan, ``E-lstm-d: A
  deep learning framework for dynamic network link prediction,'' \emph{arXiv
  preprint arXiv:1902.08329}, 2019.

\bibitem{nguyen2018continuous}
G.~H. Nguyen, J.~B. Lee, R.~A. Rossi, N.~K. Ahmed, E.~Koh, and S.~Kim,
  ``Continuous-time dynamic network embeddings,'' in \emph{Companion of the The
  Web Conference 2018 on The Web Conference 2018}.\hskip 1em plus 0.5em minus
  0.4em\relax International World Wide Web Conferences Steering Committee,
  2018, pp. 969--976.

\bibitem{zhang2018timers}
Z.~Zhang, P.~Cui, J.~Pei, X.~Wang, and W.~Zhu, ``Timers: Error-bounded svd
  restart on dynamic networks,'' in \emph{Thirty-Second AAAI Conference on
  Artificial Intelligence}, 2018.

\bibitem{zhou2018dynamic}
L.~Zhou, Y.~Yang, X.~Ren, F.~Wu, and Y.~Zhuang, ``Dynamic network embedding by
  modeling triadic closure process,'' in \emph{Thirty-Second AAAI Conference on
  Artificial Intelligence}, 2018.

\bibitem{nagaraja2010impact}
S.~Nagaraja, ``The impact of unlinkability on adversarial community detection:
  effects and countermeasures,'' in \emph{International Symposium on Privacy
  Enhancing Technologies Symposium}.\hskip 1em plus 0.5em minus 0.4em\relax
  Springer, 2010, pp. 253--272.

\bibitem{waniek2018hiding}
M.~Waniek, T.~P. Michalak, M.~J. Wooldridge, and T.~Rahwan, ``Hiding
  individuals and communities in a social network,'' \emph{Nature Human
  Behaviour}, vol.~2, no.~2, p. 139, 2018.

\bibitem{kipf2016semi}
T.~N. Kipf and M.~Welling, ``Semi-supervised classification with graph
  convolutional networks,'' \emph{arXiv preprint arXiv:1609.02907}, 2016.

\bibitem{zugner2018adversarial}
D.~Z{\"u}gner, A.~Akbarnejad, and S.~G{\"u}nnemann, ``Adversarial attacks on
  neural networks for graph data,'' in \emph{Proceedings of the 24th ACM SIGKDD
  International Conference on Knowledge Discovery \& Data Mining}.\hskip 1em
  plus 0.5em minus 0.4em\relax ACM, 2018, pp. 2847--2856.

\bibitem{chen2018fast}
J.~Chen, Y.~Wu, X.~Xu, Y.~Chen, H.~Zheng, and Q.~Xuan, ``Fast gradient attack
  on network embedding,'' \emph{arXiv preprint arXiv:1809.02797}, 2018.

\bibitem{wang2018attack}
X.~Wang, J.~Eaton, C.-J. Hsieh, and F.~Wu, ``Attack graph convolutional
  networks by adding fake nodes,'' \emph{arXiv preprint arXiv:1810.10751},
  2018.

\bibitem{sun2018data}
M.~Sun, J.~Tang, H.~Li, B.~Li, C.~Xiao, Y.~Chen, and D.~Song, ``Data poisoning
  attack against unsupervised node embedding methods,'' \emph{arXiv preprint
  arXiv:1810.12881}, 2018.

\bibitem{perozzi2014deepwalk}
B.~Perozzi, R.~Al-Rfou, and S.~Skiena, ``Deepwalk: Online learning of social
  representations,'' in \emph{Proceedings of the 20th ACM SIGKDD international
  conference on Knowledge discovery and data mining}.\hskip 1em plus 0.5em
  minus 0.4em\relax ACM, 2014, pp. 701--710.

\bibitem{tang2015line}
J.~Tang, M.~Qu, M.~Wang, M.~Zhang, J.~Yan, and Q.~Mei, ``Line: Large-scale
  information network embedding,'' in \emph{Proceedings of the 24th
  international conference on world wide web}.\hskip 1em plus 0.5em minus
  0.4em\relax International World Wide Web Conferences Steering Committee,
  2015, pp. 1067--1077.

\bibitem{konect:2017:radoslaw}
\BIBentryALTinterwordspacing
``Manufacturing emails network dataset -- {KONECT},'' Apr. 2017. [Online].
  Available: \url{http://konect.uni-koblenz.de/networks/radoslaw\_email}
\BIBentrySTDinterwordspacing

\bibitem{konect:2016:lkml-reply}
\BIBentryALTinterwordspacing
``Linux kernel mailing list replies network dataset -- {KONECT},'' Sep. 2016.
  [Online]. Available: \url{http://konect.uni-koblenz.de/networks/lkml-reply}
\BIBentrySTDinterwordspacing

\bibitem{konect:2016:dnc-temporalGraph}
\BIBentryALTinterwordspacing
``Dnc co-recipient network dataset -- {KONECT},'' Sep. 2016. [Online].
  Available: \url{http://konect.uni-koblenz.de/networks/dnc-temporalGraph}
\BIBentrySTDinterwordspacing

\bibitem{graphrepository2013}
\BIBentryALTinterwordspacing
R.~Rossi and N.~Ahmed, ``Network repository,'' 2013. [Online]. Available:
  \url{http://networkrepository.com}
\BIBentrySTDinterwordspacing

\bibitem{chen2018understanding}
T.~Chen, Y.~Zhu, Z.~Li, J.~Chen, X.~Li, X.~Luo, X.~Lin, and X.~Zhange,
  ``Understanding ethereum via graph analysis,'' in \emph{IEEE INFOCOM
  2018-IEEE Conference on Computer Communications}.\hskip 1em plus 0.5em minus
  0.4em\relax IEEE, 2018, pp. 1484--1492.

\bibitem{wu2019t}
J.~Wu, D.~Lin, Z.~Zheng, and Q.~Yuan, ``T-edge: Temporal weighted multidigraph
  embedding for ethereum transaction network analysis,'' \emph{arXiv preprint
  arXiv:1905.08038}, 2019.

\bibitem{lin2020modeling}
D.~Lin, J.~Wu, Q.~Yuan, and Z.~Zheng, ``Modeling and understanding ethereum
  transaction records via a complex network approach,'' \emph{IEEE Transactions
  on Circuits and Systems--II: Express Briefs}, 2020.

\end{thebibliography}

\end{document}